# Threshold electronic structure at the oxygen K edge of 3d transition metal oxides: a configuration interaction approach.

J. van Elp

Hiroshima Synchrotron Radiation Center

Hiroshima University, 2-313 Kagamiyama, Higashi-Hiroshima 739-8526, Japan

Arata Tanaka

Department of Quantum Matter, ADSM

Hiroshima University, 1-3-1 Kagamiyama, Higashi-Hiroshima 739-8526, Japan

**Abstract**

It has been generally accepted that the threshold structure observed in the oxygen K edge X-ray absorption spectrum in 3d transition metal oxides represents the electronic structure of the 3d transition metal. There is, however, no consensus about the correct description. We present an interpretation, which includes both ground state hybridization and electron correlation. It is based on a configuration interaction cluster calculation using a $MO_6$ cluster. The oxygen K edge spectrum is calculated by annihilating a ligand hole in the ground state and is compared to calculations representing inverse photoemission experiments in which a 3d transition metal electron is added. Clear differences are observed related to the amount of ligand hole created in the ground state. Two "rules" connected to this are discussed. Comparison with experimental data of some early transition metal compounds is made and shows that this simple cluster approach explains the experimental features quite well.

**Introduction**

Transition metal oxides have attracted attention for several decades with interest being revived in recent years because of the discovery of the Cu based high $T_c$ superconductors and the Mn based colossal magneto resistance perovskites. High energy spectroscopies like X-ray Photoemission Spectroscopy (XPS), Bremstrahlung Isochromat Spectroscopy (BIS) and X-ray Absorption Spectroscopy (XAS) at the transition metal L edges and oxygen K edges [1,2,3,4] or the Energy Electron Loss Spectroscopy (EELS) features at these edges [5,6,7] have played a key role in describing the underlying electronic structure of transition metal oxides. Experiments at the oxygen K edge have for instance determined the amount of oxygen holes upon doping in the high $T_c$ superconductors [1,2], and have recently been used to study the electronic structure of the manganese perovskites [3].

Soft X-ray absorption at the oxygen K edge is a dipole allowed transition in which an oxygen 1 s electron is promoted to an empty 2p (np) oxygen orbital. It has been generally accepted that the threshold structure observed at the oxygen K edge represents and depends on the electronic structure of the 3d transition metal ion. There is, however, no consensus about the correct description.

The 3d transition metal is important because in an ionic picture the oxygen 2p shell is full. Empty oxygen 2p orbitals are created by ground state hybridization between the 3d transition metal and oxygen 2p orbitals. After the absorption process, the extra electron is located on the 3d transition metal ion. Using a simple picture, this would suggest that oxygen K edge absorption can be interpreted in a similar way as 3d electron addition experiments like BIS.

This approach, in which energy positions deduced at the oxygen K edge are used as 3d electron addition states, was recently used to study the electronic structure of a Mn perovskite system [3]. It is partly based on work by de Groot *et al.* [8], who described the threshold structure at the oxygen K edge as resulting from a combination of ligand field splitting and exchange coupling of the 3d transition metal. The intensity or spectral weight at the oxygen K edge is directly related to the number of unoccupied 3d electrons. For the less than half filled 3d shell transition metal systems the splitting of the two threshold peaks was ascribed to ligand field splittings, but the relative intensities of the peaks could not be fully explained.

The peak positions of the empty 3d electron orbitals for the three most common oxidation states of Mn were obtained from molecular orbital theory [9] and compared by Kurata and Colliex [5] to the oxygen K edge energy electron loss spectra (EELS). The intensities were again taken as equal to the number of unoccupied electrons in each molecular orbital.

What is not explained in both of these studies is the mechanism in which the intensity is obtained. Starting from an ionic approach with full 2p orbitals, the intensity is obtained from the ground state hybridization between the 3d transition orbitals and the oxygen 2p orbitals. The strength of the hybridization is important and this has been experimentally established. Kuiper *et al.* [10] compared the oxygen K edge spectrum of $La_{2-x}Sr_xNiO_4$ directly with the high energy inverse photoemission spectrum. In the inverse photoemission data clear empty La 4f structure is present at 9 eV above the Fermi level. The intensity decreases upon Sr substitution. This La 4f structure is virtually absent in the oxygen K edge spectrum. The reason is the absence of any significant hybridization between the oxygen 2p and the very localized La 4f core valence orbitals. This means that the strength of the hybridization is important and must be included in the models used. Only counting the number of unoccupied 3d orbitals to determine the intensity is not sufficient.

Hybridization is included if one compares the oxygen K edge with the oxygen 2p projected density of states obtained from band structure calculations. For the $d^0$ compounds $SrTiO_3$ and $TiO_2$, the oxygen 2p projected density of states showed a good agreement with the oxygen K edge spectrum [11]. No influence of the oxygen core hole potential on the spectrum in the form of shifts of spectral weight was observed. Besides the $d^0$ compounds, a few oxygen K edges of $3d^n$ systems have been described using oxygen 2p projected density of states [12,13,14]. For CuO, it was found that the oxygen 2p orbitals are present up till 15 eV above the threshold and that the oxygen K edge shows structure connected to the 2p and not 3p levels up to 15 eV above the threshold [14].

What is absent in all the mentioned interpretations is electron correlation. If we accept that the threshold structure is connected to addition of 3d electrons, then starting from the ionic picture ($3d^n$) an electron from the oxygen band must be transferred to the 3d transition metal atom to obtain oxygen 2p holes ($3d^{n+1}\underline{L}$, where $\underline{L}$ stands for unoccupied oxygen orbitals). But upon adding an 3d electron to the transition metal, electron correlation is important and should be explicitly included. For inverse photoemission experiments in which the final states are mainly of $d^{n+1}$ character, multiplet structures resulting from the electron correlation are generally accepted.

In the three systems in which the oxygen K edges have been compared to the 2p projected density of states, electron correlation effects are absent. For the $d^0$ systems [11] only final states with one 3d electron can be reached and for CuO [14] the final states comprise a full 3d orbital system ($d^{10}$). In both systems the 3d electron-electron interactions are absent. $LiCoO_2$ is a special case because of the low spin $d^6$ ground state with mainly $t_{2g}^6$ character. The first electron addition state [15] is a single state reached by adding an electron of $e_g$ character, so no multiplet effects are visible in the inverse photoemission spectrum as well as at the oxygen K edge.

Configuration interaction cluster models describing the (inverse) photoemission and X-ray absorption spectra have been quite successful. Besides the hybridization, the multiplet structure in these models is included through the use of two electron operators. In the calculations in which clusters are used, the transition metal atom is mostly in the middle ($MO_6$ cluster). However for an oxygen K edge calculation one would like to have the absorbing oxygen atom in the middle surrounded by its natural environment. Such a large cluster would for a rigid octahedral transition metal monoxide cluster consist of the middle oxygen and six surrounding transition metals each with additional oxygen atoms. The total cluster would be $OM_6O_{18}$. However for the early transition metals a large number of empty 3d orbitals for every metal site is needed and the computing problem is at present not (yet) solvable. The only clusters with an oxygen in the middle used up till now are of the $Cu_2O_7$ type [16] or extensions of this [17]. The total number of holes is limited with 1 or 2 holes on every Cu atom.

Connected to this problem is also the fact that for perovskite oxides or other oxide systems, the clusters need to be more complicated and can have oxygen atoms at chemically different sites, which would require the use of different clusters. We have therefore used the relatively simple $MO_6$ cluster with oxygen molecular orbitals of $t_{2g}$ and $e_g$ symmetry. The total number of orbitals is 20. The computing problem has been solved for the interesting early transition metals with more than 5 holes in these clusters. Several studies describing the electronic spectra of valence band photoemission and inverse photoemission have been published, which gives us the opportunity to use some of these parameters.

The oxygen K edge absorption process is from a localized oxygen 1s core orbital to delocalized $e_g$ and $t_{2g}$ symmetry orbitals in our cluster. These delocalized orbitals are combinations of all six oxygens and we assume that the spectral weight to the delocalized orbitals is the same as in a calculation with the absorbing atom in the middle. The interaction of the oxygen core hole with the oxygen or transition metal valence electrons is also neglected. The bandstructure calculations showed for the $d^0$ systems [11] that the core hole has no or very little influence on the position of the final state energy levels. This is quite different from the 3d transition metal 2p edges, where a strong interaction between the 2p core hole and 3d electrons [18] is present.

A similar cluster model has been used to calculate the oxygen K edge absorption spectrum and inverse photoemission spectrum of $LiNiO_2$ [19]. The trivalent Ni groundstate is low spin, so addition of spin up and spin down electrons is possible. The character of the ground state is mainly $d^7\underline{L}$. The configuration interaction cluster calculations describes the spectral weight difference between the threshold structure at the oxygen K edge and first part of the inverse photoemission spectrum quite well. The present results are an extension of those calculations to $3d^1$ to $3d^7$ transition metal systems.

**Calculational details and parameters**

The cluster used is an octahedral transition metal symmetry $MO_6$ cluster and is similar to clusters used in the analysis of late transition metal high energy spectroscopies [15,19,20]. It has 10 3d orbitals and 10 oxygen molecular orbitals of $t_{2g}$ and $e_g$ symmetry. The non bonding oxygen orbitals are omitted. The d-d Coulomb and exchange interactions are included using atomic multiplet theory specified in terms of Racah A, B, and C parameters. If available, the free ion values of the Racah B and C parameters, as tabulated by Griffith [21], are used. For the $Mn^{4+}$ system scaled down values (80%) of calculated Hartree-Fock values are used [18]. The $d^1$ system $Ti^{3+}$ has no electron correlation. Estimated values based on $Ti^{2+}$ were used. All the parameters are listed in Table 1.

The Mott-Hubbard energy U is defined as the energy needed to remove a 3d transition metal electron and add it at another $d^n$ site and describes the transition from $d^n d^n \rightarrow d^{n+1} d^{n-1}$. The lowest energy level of each $d^n$ configuration is calculated with the ionic part of the transition metal ligand field splitting ($10Dq^i$) zero.

The charge transfer energy $\Delta$ is defined as the energy needed for transferring an electron from the lowest level of the $d^n$ configuration to the lowest level of the $d^{n+1}\underline{L}$ configuration. The pp$\sigma$, pp$\pi$ and $10Dq^i$ parameters are zero in this definition. The hybridization between the transition metal 3d states and the ligand 2p orbitals is taken into account by the Slater-Koster [22] pd$\sigma$ and pd$\pi$ transfer integrals. they describe the overlap between the oxygen and transition metal orbitals for $\sigma$ and $\pi$ bonding. According to Harrison [23], the value of pd$\pi$ is fixed at pd$\pi$=-0.45pd$\sigma$.

The Slater-Koster [22] oxygen nearest neighbor interactions pp$\sigma$ and pp$\pi$ split the oxygen states in a double degenerate level with $e_g$ symmetry at pp$\sigma$-pp$\pi$ and a triple degenerate level with $t_{2g}$ symmetry at -(pp$\sigma$-pp$\pi$). The width of the oxygen band determines the value of pp$\sigma$-pp$\pi$.

In the ground state, charge transfer levels up to $d^{n+2}\underline{L}^2$ are included, while in the final states only one charge transfer level ($d^{n+2}\underline{L}$) is included. In the final state the energy difference between $d^{n+1}$ and $d^{n+2}\underline{L}$ is U+$\Delta$, which means that double charge transfer states are at very high energy and can be neglected.

|  | A | B | C | $\Delta$ | pd$\sigma$ | 10Dq | pp$\sigma$-pp$\pi$ |
|---|---|---|---|---|---|---|---|
| $Co^{2+}$ | 5.2 | 0.138 | 0.541 | 5.5 | 1.3 | 0.7 | 0.7 |
| $Fe^{2+}$ | 5.5 | 0.131 | 0.483 | 7.0 | 1.3 | 0.7 | 0.7 |
| $Mn^{2+}$ | 3.9 | 0.119 | 0.412 | 8.8 | 1.3 | 0.7 | 0.7 |
| $Fe^{3+}$ | 4 | 0.126 | 0.595 | 5 | 1.9 | 1.3 | 0.8 |
| $Mn^{3+}$ | 5 | 0.141 | 0.456 | 5 | 2.1 | 1.3 | 0.8 |
| $Cr^{3+}$ | 5 | 0.128 | 0.477 | 5 | 2.3 | 1.3 | 0.8 |
| $Mn^{4+}$ | 5 | 0.132 | 0.497 | 4 | 2.5 | 1.5 | 0.8 |
| $V^{3+}$ | 5 | 0.106 | 0.516 | 5 | 2.3 | 1.3 | 0.8 |
| $Ti^{3+}$ | 5 | 0.1* | 0.5* | 5 | 2.3 | 1.3 | 0.8 |

*Table 1. Parameters used for the calculations, all values in eV. The Racah B and C parameters for trivalent Ti are estimates as the $d^1$ system has no two electron interactions.*

Probably the most important parameter is the ionic $10Dq^i$ splitting. It splits the 3d orbitals in $e_g$ and $t_{2g}$ orbitals at $+6Dq$ and $-4Dq$, respectively. This splitting has not always been included in the analysis of valence band photoemission spectra because the splittings it induces in the electron removal final states are rather small as compared to the valence band width. For NiO and CuO the valance band widths are about 12 eV and for both a good result [24,25] can be obtained without the use of an ionic $10Dq^i$. The main reason is that the overall width is obtained from hybridization between almost degenerate $d^{n-1}$ and $d^n\underline{L}$ final states.

If the same parameters are to describe the forbidden 3d-3d optical transitions, an ionic $10Dq^i$ splitting must be added [19]. This is most apparent for the $d^3$ system. The experimentally determined 10Dq splitting between the $^4A_{2g}$ ground state and $^4T_{2g}$ excited state is about 2 eV for $Cr^{3+}$ [26] and is independent of Racah B and C parameters. Neglecting the ionic $10Dq^i$ contribution, one finds a calculated splitting of less than 1 eV using the $MO_6$ cluster. The splitting is caused by a hybridization between $d^n$ and $d^{n+1}\underline{L}$ configurations through the difference in transfer integrals ($pd\sigma$, $pd\pi$) for $e_g$ and $t_{2g}$ electrons. For the $Cr^{3+}$ system an ionic $10Dq^i$ of about 1.3 eV needs to be added to obtain the correct experimental 10Dq splitting of 2 eV between the $^4A_{2g}$ and $^4T_{2g}$ states.

The ionic $10Dq^i$ value was determined for the $Fe^{3+}$ system using published optical transitions [27] of $Fe^{3+}$ impurities in $Al_2O_3$. In the $Cr^{3+}$ system the optical 10Dq value is independent of Racah B and C parameters, while for the $Fe^{3+}$ system these parameters also determine the optical splittings. For both systems an ionic $10Dq^i$ of 1.3 eV [28] is obtained and this values was also used for the $d^1$, $d^2$ and $d^4$ systems. A detailed analysis of the optical transitions using configuration interaction cluster calculations will be presented in the near future [28].

The full 10Dq splitting is different for different final states and is hybridization dependent. For MnO [29] the optical 10Dq is 1.2 eV with hybridization between $d^5$ and $d^6\underline{L}$, which have an energy difference determined by the charge transfer energy $\Delta$. For the $^5T_{2g}$-$^5E_g$ splitting in photoemission and inverse photoemission spectra [20] values of 1.9 eV ($d^4$ and $d^5\underline{L}$, $\Delta$-U apart) and 1.0 eV ($d^6$ and $d^7\underline{L}$, $\Delta$+U apart) are obtained. The stronger the hybridization, because of the proximity of the levels involved, the larger is the splitting.

The values for the trivalent systems of the transfer integral $pd\sigma$ is obtained by increasing the integrals from the values obtained for the divalent Mn and Fe systems. Values for $pd\sigma$, based on fitted tight-binding parameters to a linearized muffin-tin orbital band structure calculation for $LaMO_3$ perovskites systems (M ranges from Ti to Ni) [30], are for the early transition metals around 2.3 eV. This value was chosen for trivalent Ti to Cr systems, while for the trivalent Mn [30] system 2.1 eV was used. For the trivalent Fe system a slightly larger value as compared to the $LaMO_3$ systems is deduced.

The optical 10Dq transition in $Mn^{4+}$ doped $Al_2O_3$ is around 2.4 eV [31]. Besides the larger 10Dq splitting a larger covalency is also expected, so the ionic $10Dq^i$ is increased to 1.5 eV and $pd\sigma$ to 2.5 eV and the charge transfer energy $\Delta$ for $Mn^{4+}$ systems is decreased to 4 eV.

The parameters for the $Mn^{2+}$, $Fe^{2+}$ and $Co^{2+}$ systems are obtained from an analysis of combined photoemission and inverse photoemission experiments using similar configuration interaction cluster calculations [15,20,32]. The obtained values for Racah A and $\Delta$ are used. The parameters obtained from an analysis of the 2p resonant photoemission spectra are quite similar [33].

The inverse photoemission or BIS is calculated using the addition of a 3d electron. Instead of electron addition, a ligand hole of oxygen character is annihilated in the oxygen K edge calculations. Both processes reach the same final states, $d^{n+1}$ and $d^{n+2}\underline{L}$, but the branching ratios are different. Annihilating a ligand hole in our cluster describes the transition at the oxygen edge as an $d^{n+1}\underline{L} \rightarrow \underline{c}d^{n+1}$ (where $\underline{c}$ stands for the oxygen core hole). The intensity of the oxygen K edge absorption spectrum $I_{O1s}(\omega)$, which is a function of the incident photon energy $\omega$, is formally given by:

$$I_{O1s}(\omega) = \sum_{f,\upsilon,\sigma} \left|\langle g|l_{\upsilon,\sigma}|f\rangle\right|^2 \delta(E_f - \hbar\omega - E_g)$$

The $E_g$ and $|g\rangle$ denote the eigenvalue and the eigenfunction of the ground state and $E_f$ and $|f\rangle$, those of the final states. The opearator $l_{\upsilon,\sigma}$ annihilates a hole with spin quantum number $\sigma$ at the ligand orbital labeled $\upsilon$. The core hole and its interactions with the oxygen or transition metal atoms are ignored in the calculations. Note that our results are identical to those obtained from a $MO_6$ cluster calculation with all O 2p orbitals included, if we neglect the O 2p on site Coulomb interaction and the O 1s core hole potential.

### Results

The calculations are divided into two groups. In Fig. 1 the BIS and oxygen K edge spectra of systems with five or more 3d electrons are shown. Here the number of finals states is limited because the only electron addition possible is always a minority spin electron. For $d^8$ in $O_h$ symmetry (NiO) and $d^9$ systems (CuO and the parent systems of the high $T_c$ superconductors), the number of possible final states is limited to one. From $d^4$ downwards, the number of final states changes and a much larger number can be reached (see Fig. 2). Besides a minority spin electron, now the possibility of adding a majority spin electron also exists.

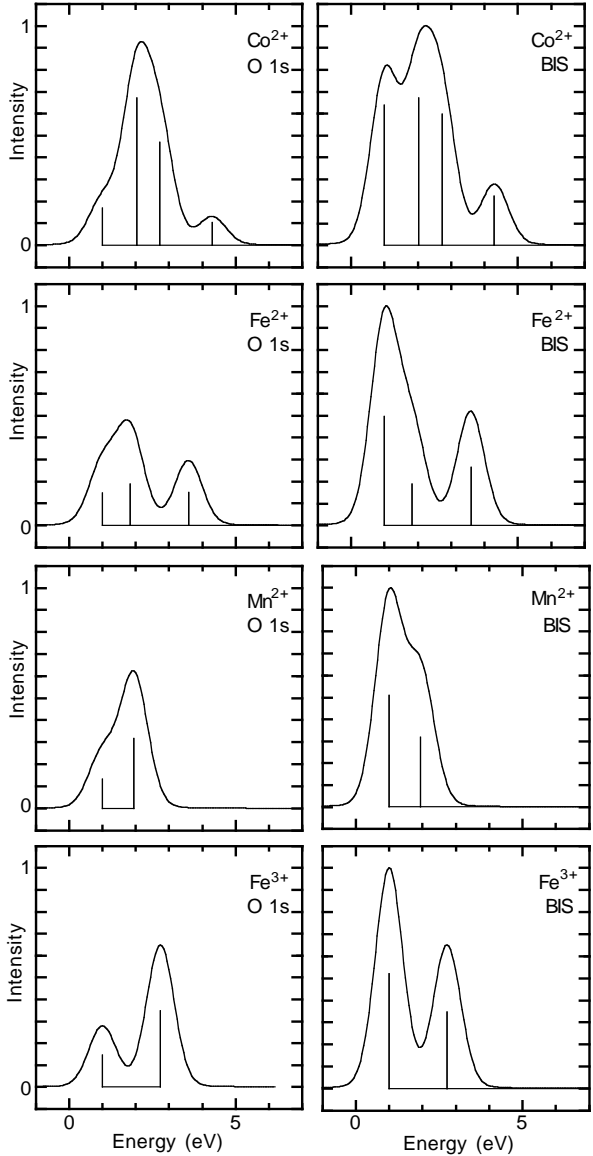

Fig. 1. Oxygen K edge and inverse photoemission (BIS) calculations for the $d^7$ to $d^5$ systems $Co^{2+}$, $Fe^{2+}$, $Mn^{2+}$ and $Fe^{3+}$. A standard broadening of 0.8 eV is used. The sticks show the strength of individual final states.

The scaling between the two calculations is such that the least effected peak in the oxygen K edge calculation (O 1s) has the same strength as in the BIS calculation. This is almost always the lowest energy final state where an $e_g$ electron is added. For convenience, the lowest enery final states in the figures are aligned to each other at 1 eV. It is clear that the intensity of some final states changes significantly between the two types of calculations.

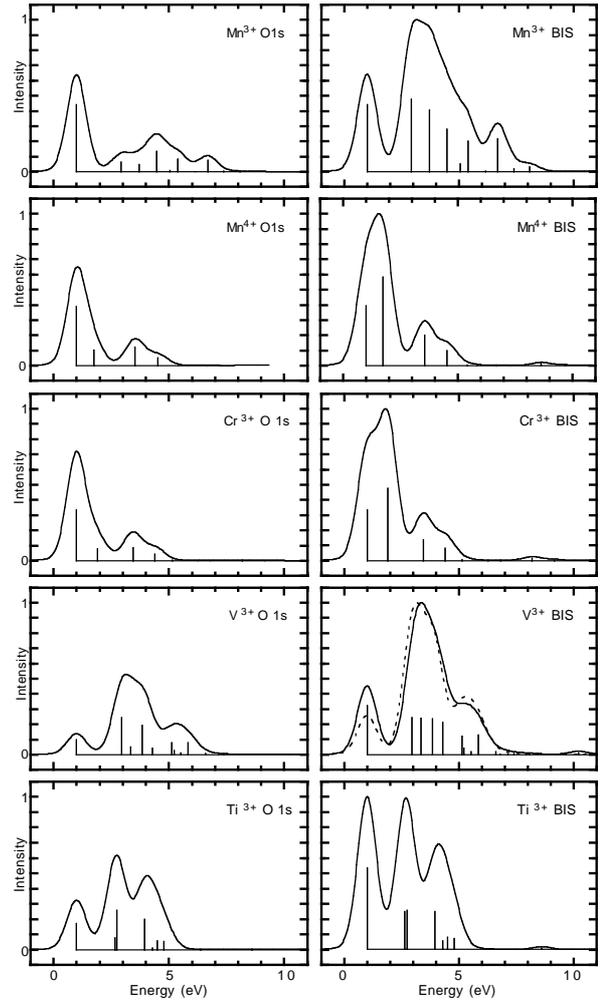

Fig. 2. Oxygen K edge and inverse photoemission (BIS) calculations for $d^4$ to $d^1$ systems. A standard broadening of 0.8 eV is used. The sticks show the strength of individual final states.

The positions of the final states reached for $n \geq 5$ are similar to the positions in the optical spectrum of the $d^{n+1}$ system, in which only spin allowed transitions are observed [34]. The spin flip transitions are not accessible. For the $Co^{2+}$, the final states obtained are of $^3A_{2g}$, $^3T_{2g}$, $^3T_{1g}$ and $^3T_{1g}$ symmetry, and for the $d^5$ systems only two final states of $^5T_{2g}$ and $^5E_g$ symmetry are obtained. In the $Fe^{2+}$, there are 3 instead of the expected 4 final states with $^4T_{1g}$, $^4T_{2g}$ and $^4T_{1g}$ symmetry. The $^4A_{2g}$ cannot be reached with the addition of an $e_g$ or $t_{2g}$ electron from a $^5T_{2g}$ ground state.

For the $n<5$ systems, the number of final states is much bigger. The lowest energy electron addition state is the ground state for the $n+1$ system. The lowest energy electron addition state is formed by adding an electron determined by Hund's rule. It has $^3T_{1g}$ symmetry for $d^1$, $^4A_{2g}$ for $d^2$, both with the addition of a majority spin $t_{2g}$ electron in the first fi-nal state. The $^5E_g$ symmetry for $d^3$ and $^6A_{1g}$ symmetry for $d^4$ are reached both with the addition of a majority spin $e_g$ electron. The oxygen ligands in transition metal oxides are not strong enough to create low spin ground states. The lowest energy electron addition state is of the same character as based on ligand field and exchange arguments.

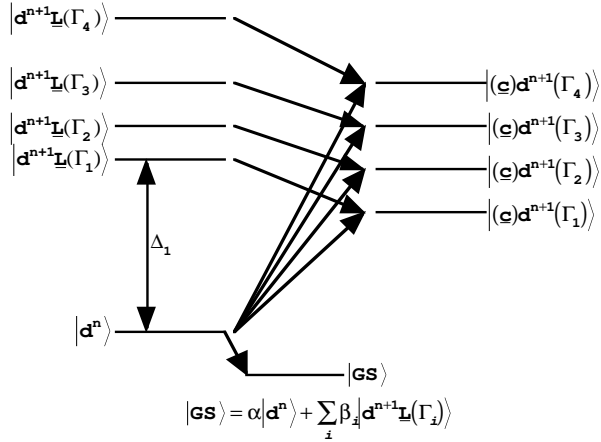

*Fig. 3. Schematic picture of the difference between a transition at the oxygen K edge and an electron addition transition as measured using BIS.*

Fig. 3 shows the qualitative difference between 3d electron addition and oxygen hole annihilation. Simplified, the electron addition process is $d^n \rightarrow d^{n+1}$. The individual strength of the final states is determined by the Wigner coefficients connected to each individual final state transition. The annihilation of an oxygen hole can be described in a simplified picture as $d^{n+1}\underline{L} \rightarrow \underline{c}d^{n+1}$, omitting the core hole. The transition is straightforward and the strength is directly connected to the amount of ligand hole present in the ground state. In mixing the $d^{n+1}\underline{L}$ states into the ground state the Wigner coefficients are of course involved and determine for a part the amount of mixing possible for each $d^{n+1}\underline{L}$ state. But besides these coefficients, the amount of overlap or the size of the transfer integrals is important. To understand their influence on the oxygen K edge spectrum simple perturbation theory arguments can be used.

In first order perturbation theory, the amount of $d^{n+1}\underline{L}$ mixed into the $d^n$ ground state wave function is equal to $T/\Delta$, with T the transfer integrals or overlap integral, and $\Delta$ the energy difference between the ground state and the perturbing level. The amount of ligand hole created in this way is equal to $(T/\Delta)^2$. From this, it is directly possible to deduce two important rules for the interpretation of the oxygen K edges threshold electronic structure.

First, the ligand hole created upon the transfer of a $e_g$ electron results in an intensity in the spectrum that is four times stronger than for $t_{2g}$ electrons, as compared to the $d^n$ electron addition spectra. The transfer integrals for $e_g$ electrons are about 2 times as strong as for $t_{2g}$ electrons ($pd\pi=-0.45pd\sigma$), which means that the actual ligand hole amount is about four times as large. Second, charge transfer states, which are higher in energy, have less intensity in the oxygen K edge spectrum, because the amount of ligand hole created in the ground state is smaller. This depends strongly on the charge transfer energy $\Delta$. If $\Delta$ is small, the differences in energy within the $d^{n+1}\underline{L}$ charge transfer states are large and the first peak is more pronounced. For large $\Delta$, this effect will hardly be noticeable.

The effect of both rules can be seen in almost all the systems. The first electron addition state for the $d^1$, $d^2$ and the $d^n$, $n \geq 5$ are all reached by the addition of a $t_{2g}$ electron. In all these systems, the first electron addition state has a significantly lower intensity at the oxygen K edge as compared to the BIS spectrum. For the $d^4$ system $Mn^{3+}$, the first electron addition state of $^6A_{1g}$ symmetry is the only possible way of adding an $e_g$ majority spin electron. The first minority spin $e_g$ electron addition state is at 4.5 eV and it has lost intensity as compared to the $^6A_{1g}$ state, because it's charge transfer energy $\Delta$ is much larger. For the $d^3$ and $d^4$ systems, the oxygen K edge spectrum is almost fully made up of final states connected to the addition of a $e_g$ electron. The differences between the BIS and oxygen K edge are large. For $d^1$ and $d^2$, the main difference is the first peak because the overall distribution of $e_g$ and $t_{2g}$ electron addition final states is similar. This can be seen in the $V^{3+}$ BIS spectrum where also the oxygen K edge spectrum has been plotted (see Fig. 2) normalized to the same maximum. Overall, the effects of both "rules" are clearly visible.

The energy positions at the oxygen K edges are not very sensitive to small changes in parameters except for the ionic $10Dq^i$ contribution. This parameter determines together with the Racah B and C parameters mainly the splittings. A change from $10Dq^i=0.7$ eV for the $Mn^{2+}$ to 1.3 eV for $Fe^{3+}$ increases the $^5T_{2g}$ to $^5E_g$ splitting from 0.9 eV to 1.7 eV. The increase from the larger transfer integrals of $Fe^{3+}$ is limited. The Racah B and C are fixed parameters, values chosen based on the free atom values for each individual ion. Changes in the oxygen bandwidth factorized in $pp\sigma$-$pp\pi$ do not effect the overall spectrum much. The electron removal spectrum in cluster calculations is very sensitive to the difference between U (through Racah A) and $\Delta$. This is not the case for electron addition, where the first possible energy position for hybridization in the final states is at U+Δ. Compared to the amount of $d^n$ character in the ground states, the final states have a larger amount of $d^{n+1}$ character .

**Comparison with experimental systems**

In Fig. 4, the calculated results for the oxygen K edge are compared with experimental spectra obtained from the literature for $d^3$ (LiMnO$_3$) [13], $d^2$ (LiVO$_2$) [35] and $d^1$ systems (Ti$_2$O$_3$ [8] and LaTiO$_3$ [36]). In transition metal systems with n<5, the majority spin band is not full. In addition to minority spins electrons, also majority spin electrons can be added. The possible number of final states is much larger and the width of the final state multiplet spectrum is much larger too, and therefore a more meaningful comparison with experimental systems can be made.

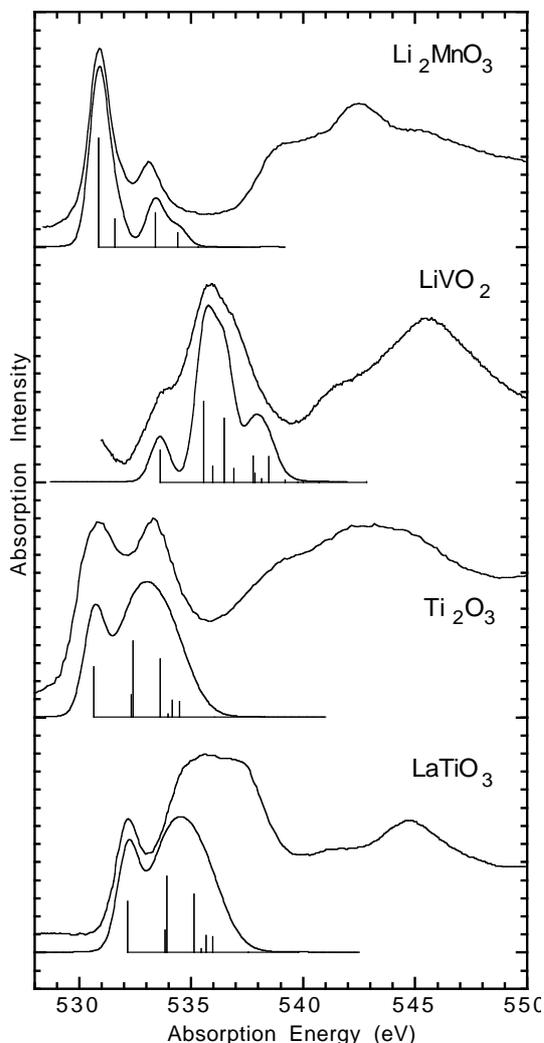

*Fig. 4. Comparison between the experimental oxygen K edges and the calculations for the $d^1$, $d^2$, $d^3$ systems LaTiO$_3$, Ti$_2$O$_3$, LiVO$_2$ and LiMnO$_3$. In the $d^1$ calculation the final states beyond 1 eV are broadened with 1.7 eV.*

The features at threshold of the LiMnO$_3$ oxygen K edge spectrum are well reproduced in the calculation. In the original work by de Groot *et al.* [8] and in the molecular orbit calculations [5,9], the interpretation for $d^3$ systems is different. The first peak, which in our case is the addition of a majority spin $e_g$ electron, is explained as a combination of majority spin $e_g$ and minority spin $t_{2g}$ bands with relative intensity of 40% and 60%, respectively. intensities equal to the number of empty 3d orbitals. The second peak, which in our case mainly consists of minority spin $e_g$ electron addition, split by multiplet structure, is now an only minority spin $e_g$ band. Because of hybridization, the $t_{2g}$ electron addition shows up weak in the cluster calculations and is positioned in between the two $e_g$ related structures.

The electronic structure of LiVO$_2$, including the phase transition at 460 K, has been discussed extensively by Pen *et al.* [35]. It is clear that the overall shape of the oxygen K edge spectrum is reproduced quite well. The first peak is electron addition of a majority spin $t_{2g}$ electron. The $e_g$ bandwidth is large and the $t_{2g}$ structure is only visible as a shoulder of the $e_g$ derived structures at 536 eV.

In both the LiVO$_2$ [37] and the LiMnO$_3$ [38] systems, the transition metal ion is in a close to octahedral symmetry. In both systems the Li plays no role in the electronic structure. This transition metal octahedral symmetry is not observed in Ti$_2$O$_3$. The TiO$_6$ octahedra are distorted along the trigonal axis [39] and three metal oxygen bond distances are 2.08 Å and three are 2.01 Å. Besides the distortion also the Ti atoms have one Ti neighbor with a short metal to metal distance which points to possible metal to metal interactions as proposed by Goodenough [40]. The trigonal distortion splits the $t_{2g}$ orbitals into an $a_1$ and double degenerate e symmetry orbital. The $e_g$ orbitals do not split but change symmetry to e too, and hybridize with the e symmetry orbitals derived from the $t_{2g}$ orbitals. The splitting of the $t_{2g}$ orbitals is only important for the $t_{2g}$ derived first final state, which will obtain an extra broadening.

More important, however, is that the distortion, which also changes the bond angles of the octahedron, mixes the $e_g$ (pdσ) and $t_{2g}$ (pdπ) transfer integrals. The amount of ligand hole created for $e_g$ or $t_{2g}$ charge transfer changes. This means that for the Ti calculation, the $t_{2g}$ derived first peak obtains more intensity. This effect is already visible at the oxygen K edge of Ti$_2$O$_3$, combined with the fact that the $e_g$ electron addition final states have a much larger bandwidth. Overall, the width observed at the threshold is comparable to the width of the multiplet structure calculated although the exact intensities are not fully reproduced.

The last example to discuss is LaTiO$_3$. In addition to the Ti, also the La empty orbitals can hybridize with oxygen orbitals as is evident from the structure observed at 535 eV, which is not only connected to Ti but also to La empty orbitals. The first peak at the oxygen K edge can clearly be identified as a Ti empty orbital with majority spin t$_{2g}$ character only. In this system we also have chemically different oxygen atoms, one in the LaO layer and one in the TiO$_2$ layer.

**Discussion**

In the previous section, we compared our calculated results with experimental systems with a less than half filled 3d shell. The d$^2$ and d$^3$ systems show quite clearly that at the oxygen K edge we observe electronic structure described by a combination of electron correlation, ionic contribution to a ligand field splitting and, most importantly, by hybridization between the oxygen and transition metal atom in the ground state. The actual amount of ligand holes created in the ground state is instrumental for a correct understanding of the experimental features at the oxygen K edges.

In the more than half filled systems such a clear cut situation is not observed. The different interpretations cause differences in strength of the final states and in energy positions between our approach and results based on ligand field splittings [8] or oxygen 2p partial density of states obtained from bandstructure calculations. The widths of the calculated oxygen K edge spectra and of the calculated BIS spectra are more or less similar for all the different interpretations and a limited number of features is observed in the experimental spectra [15]. Overall, it is difficult to judge and give full credit to a certain interpretation for d$^n$, n≥5 systems.

The expected strength connected to the e$_g$ electrons final states is not always experimentally observed. In LaFeO$_3$ [41] the second peak connected to the addition of an e$_g$ electron has only slightly more intensity than the first peak connected to a t$_{2g}$ orbital. Both of the e$_g$ electrons are pointing towards the oxygen neighbors and because of this, they have a much larger bandwidth. This means that if the e$_g$ bandwidth is much larger than the corresponding t$_{2g}$ bandwidth the relative intensity of the e$_g$ peaks will go down and also e$_g$ peaks can overlap the t$_{2g}$ peaks. This overlap is likely present in MnO [8] and CoO [15] as no clear t$_{2g}$ structure is visible in these two systems. In both oxygen K edge spectra the rise at threshold is much slower than what should be expected based on the resolution used. The same situation is observed for the LiVO$_2$ system, here the first structure connected to t$_{2g}$ orbitals is only present as a shoulder on the wide e$_g$ related structure.

The first order pertubation theory approach shows the extent of the influence of the hybridization on the strength of the oxygen K edge spectrum as compared to the inverse photoemission spectrum. It is important to realize that both "rules" are independent of the chosen cluster approach. They are quite similar for a single oxygen interacting with a transition metal ion, because they are obtained from a simple pertubation approach. The fact that the calculations work so well for the LiVO$_2$ and LiMnO$_3$ systems is due to the two simple "rules".

The oxygen K edge is sensitive to changes in local transition metal geometry because of the strong dependence on the hybridization. A small distortion, which in the BIS spectrum will split a peak with a few hundred meV, and which has little influence on the overall spectrum, because of a large bandwidth, has on the oxygen K edge a large effect. The size of the transfer integrals changes because the σ and π overlap changes. An angle distortion mixes them and this mixing changes the intensity to a much greater extent than what would be observed in the corresponding BIS spectrum. The oxygen K edge intensity is also not normalized to the amount of empty 3d orbitals as a BIS experiment is.

As compared to our calculation, the local Ti geometry in Ti$_2$O$_3$ will increase the intensity of the first features of t$_{2g}$ character in the oxygen K edge spectrum. The overall width is obtained but the intensities are not fully reproduced. The LaTiO$_3$ spectrum also shows the difficulty in understanding the oxygen threshold structure because both metals are hybridizing with oxygen atoms. The first peak can be identified as a majority spin t$_{2g}$ orbital or band.

The local geometry also plays a major role in the Jahn-Teller d$^4$ systems. The two atoms on the z axis have an elongated or shortened distance to the transition metal ion. The degeneracy of the e$_g$ electrons is lifted and depending on the distortion the occupied orbital is the d$_{x^2-y^2}$ (shortened) or the d$_{z^2}$ (elongated). Two chemically different oxygen atoms are created. The lowering of the symmetry leads to extra ligand field parameters [34], Ds and Dt, which describe the splitting between the two e$_g$ and between the t$_{2g}$ electrons. We have not attempted to compare our calculations with a d$^4$ system, mainly because of the following reasons.

The hybridization with the z axis oxygen atoms changes with the distortion. Depending on the out of plane to in plane bond distance this can be described with transfer integrals connected to the z axis and to the in plane xy direction. The overlap integral [23] changes with the distance ratio to the power 3.5. For a 15% elongation this change is 63%, which means $T^2$ is about a factor 2.6 larger. For an elongated system, the strongest hybridized orbital would then be the $d_{x^2-y^2}$. This would be the first $e_g$ derived final state in our calculations, reached upon filling the majority spin $d_{x^2-y^2}$. The strongest intensity minority spin orbital would also be the $d_{x^2-y^2}$. The main problem would be the position of the energy level of these orbitals because one can also expect an ionic contriution to the splitting of the $e_g$ and the splitting of the $t_{2g}$ orbitals. The splitting between the two $e_g$ and between the $t_{2g}$ orbitals can be quite large. In $CrF_2$, a splitting of 3.1 eV between the two split $e_g$ electrons is observed [42], but this system also has a large bond distance ratio of 1.22. For trivalent Mn systems values for the two split $e_g$ orbitals of about 1 eV have been observed [43]. This splitting is not only caused by differences in hybridization but also an ionic contirbution is needed. But it is not clear how large this ionic contribution to the splitting between the $e_g$ ($t_{2g}$) orbitals is. Such large distortion needs a different cluster set up where the hybridization with in plane and out of plane oxygens is properly taken into account.

The last point to mention is possible 3d transition metal threshold structures for other ligands than oxygen. One would expect to see the same spectra at the fluoride K edge (~790 eV) although with significantly smaller intensity because of a much smaller ground state hybridization between the fluoride 2p and transition metal 3d orbital.

## Conclusions

It has been generally accepted that the threshold structure observed in the oxygen K edge X-ray absorption spectrum represents the electronic structure of 3d transition metal orbitals in 3d transition metal oxides. The interpretation presented by us includes both ground state hybridization including the differences between σ bonds and π bonds and electron correlation. It is based on a configuration interaction cluster calculation using a $MO_6$ cluster. The O K edge spectrum is calculated by annihilating a ligand hole in the ground state and is compared to electron addition calculations representing inverse photoemission experiments. Clear differences are observed related to the amount of ligand hole created in the ground state.

Based on the difference and using simple perturbation theory arguments, two "rules" can be formulated. First, the ligand holes created with the charge transfer of an $e_g$ electron have four times more intensity as ligand holes created with charge transfer of a $t_{2g}$ electron. The difference is directly connected to the difference in overlap integrals (pdσ and pdπ), which reflects the difference between σ and π bonding. Second, the higher the energy of a particular charge transfer level Δ in the ground state, the weaker the intensity of this level in the oxygen K edge spectrum. Both "rules" can explain the differences between the presented oxygen K edge calculations and the electron addition calculations representing inverse photoemission spectra.

The comparison of the $d^2$ and $d^3$ calculations with the oxygen K edge data of $LiVO_2$ and $LiMnO_3$ shows that the oxygen K edge spectral weight is described quite well by taking into account the amount of ligand holes in the ground state and the electron correlation of the 3d electrons. For systems with a less than half filled shell, electron addition produces the largest energy spread in the final states because of the possibility of majority and minority spin electron addition. For the $d^3$ and $d^4$ systems the oxygen K edge spectral weight is almost completely made up of $e_g$ electron addition final states.

The comparison with the two $d^1$ systems $Ti_2O_3$ and $LaTiO_3$ also shows the limitations of our approach. Distortions of octahedral symmetry are expected to have a large influence on the oxygen K edge spectrum. Mixing of the transfer integrals occurs because of mixing of σ and π bonding between the transition metal atom and oxygen atom. In $LaTiO_3$, the hybridization of oxygen orbitals with empty La orbitals is possible and leads to structure at the oxygen K edge overlapping the 3d transition metal threshold structures.

The change of transfer integrals with distortions can have a strong effect on z axis elongated $d^4$ Jahn-Teller systems. With a large distortion, the oxygen K edge spectrum is mainly determined by final states connected to the $d_{x^2-y^2}$ orbital. The other orbitals have much smaller spectral weights. Also, such a system has two inequivalent oxygen positions, which means that in calculations larger clusters are needed.

## Acknowledgment

We thank H. Pen for sending the $LiVO_2$ spectrum in digital form to us and Prof. T. Jo for a critical reading of the manuscript. This work is partly supported by a Grant-in-Aid for scientific research from the Ministry of Education, Science, Sports and Culture of Japan.